\documentclass[usegraphicx,usenatbib]{mn2e}
\usepackage{times}
\usepackage{epsf}

\newcommand{\source}{\hbox{NGC\,315}}

\newcommand{\chandra}{\textit{Chandra}}
\newcommand{\rosat}{\textit{ROSAT}}

\title[The inner jet of \source]
{The inner jet of radio galaxy \source\ as observed with {\it
Chandra\/} and the VLA
}

\author[D.M. Worrall et al.]
 {D.M. Worrall,$^1$ M. Birkinshaw,$^1$  R.A. Laing,$^2$
 W.D. Cotton,$^3$ A.H. Bridle,$^3$\\
$^1$Department of Physics, University of Bristol, Tyndall Avenue,
Bristol BS8~1TL \\
$^2$European Southern Observatory, Karl-Schwarzschild-Stra{\ss}e 2, D-85748
       Garching-bei-Muenchen, Germany  \\
$^3$National Radio Astronomy Observatory, 520 Edgemont Road,
       Charlottesville, VA 22903-2475, U.S.A. 
 }

\begin{document}

\label{firstpage}

\maketitle

\begin{abstract}

We present \chandra\ X-ray results for the jet, nucleus, and gaseous
atmosphere of \source, a nearby radio galaxy whose jet kinematics are
known through deep radio mapping. Diffuse X-ray synchrotron emission
is detected from the jet out to 30 arcsec from the nucleus, through
regions both of fast bulk flow and deceleration.  The X-ray to radio
flux ratio drops considerably where the flow decelerates, but the
X-ray and radio emissions show similar transverse extents throughout,
requiring distributed particle acceleration to maintain the supply of
X-ray-emitting electrons.  A remarkable knotty filament within the jet
is seen in both the radio and X-ray, contributing roughly 10 per cent
of the diffuse emission along its extent at both wavelengths.  No
completely satisfactory explanation for the filament is found, though its
oscillatory appearance, roughly aligned magnetic field, and
requirements for particle acceleration, suggest that it is a magnetic
strand within a shear layer between fast inner and slower outer
flow. 

\end{abstract}


\section{Introduction}
\label{sec:intro}

The nearby elliptical galaxy \source\ has long been known to be the
host of an FRI \citep{fr} radio source whose jets extend for a degree
on the sky \citep{fanti76, bridle76, willis81, jagers87, laing06a}.
The radio jets are one-sided in their inner regions \citep[with evidence for
acceleration on parsec scales found by][]{cotton99}, but become
increasingly symmetrical on kpc scales.  This is consistent with the
hypothesis of a decelerating relativistic flow, as discussed by
\citet{bick2}.  More recent modeling by \citet{canvin} finds an
on-axis flow speed of 0.9$c$ prior to deceleration at a projected
angle of 14~arcsec from the core, and a jet inclination to the line of
sight of $\theta = 38 \pm 2$ degrees.  Deceleration is likely to
result from mass-loading of the jets, either from entrainment of the
external interstellar medium (ISM) or from stellar mass loss within
the jet. The ISM fundamentally affects jet propagation via pressure
gradients and buoyancy forces \citep[e.g.,][]{bick2, wbcn326,
laing02b}, so a knowledge of the density and pressure of the hot ISM,
which can be determined directly from X-ray observations, is essential
to an understanding of jet dynamics.

Earlier \chandra\ observations of \source\ found an X-ray jet
coincident with the stronger radio jet \citep*{wbhn315}.  The
present paper gives the results of a deeper \chandra\ observation, and
in particular discusses the association of coherent, filamentary
structures in the radio jet with X-ray emission.  We also present the
density and pressure of the ISM on the scales of the X-ray-emitting
jet, as determined from this more sensitive observation.

The redshift of \source\ is $0.01648\pm 0.00002$ \citep{trager}.
In this paper we adopt values for the cosmological parameters of $H_0
= 70$~km s$^{-1}$ Mpc$^{-1}$, $\Omega_{\rm {m0}} = 0.3$, and
$\Omega_{\Lambda 0} = 0.7$.  Thus 1~arcsec corresponds to a projected
distance of 335~pc at \source, and a deprojected distance along the jet of 544~pc.
Spectral index, $\alpha$, is defined in the sense that flux density is
proportional to $\nu^{-\alpha}$.  J2000 coordinates are used
throughout.

\section{Observations}
\label{sec:obs}

\begin{figure*}
\centering
\includegraphics[width=6.5truein]{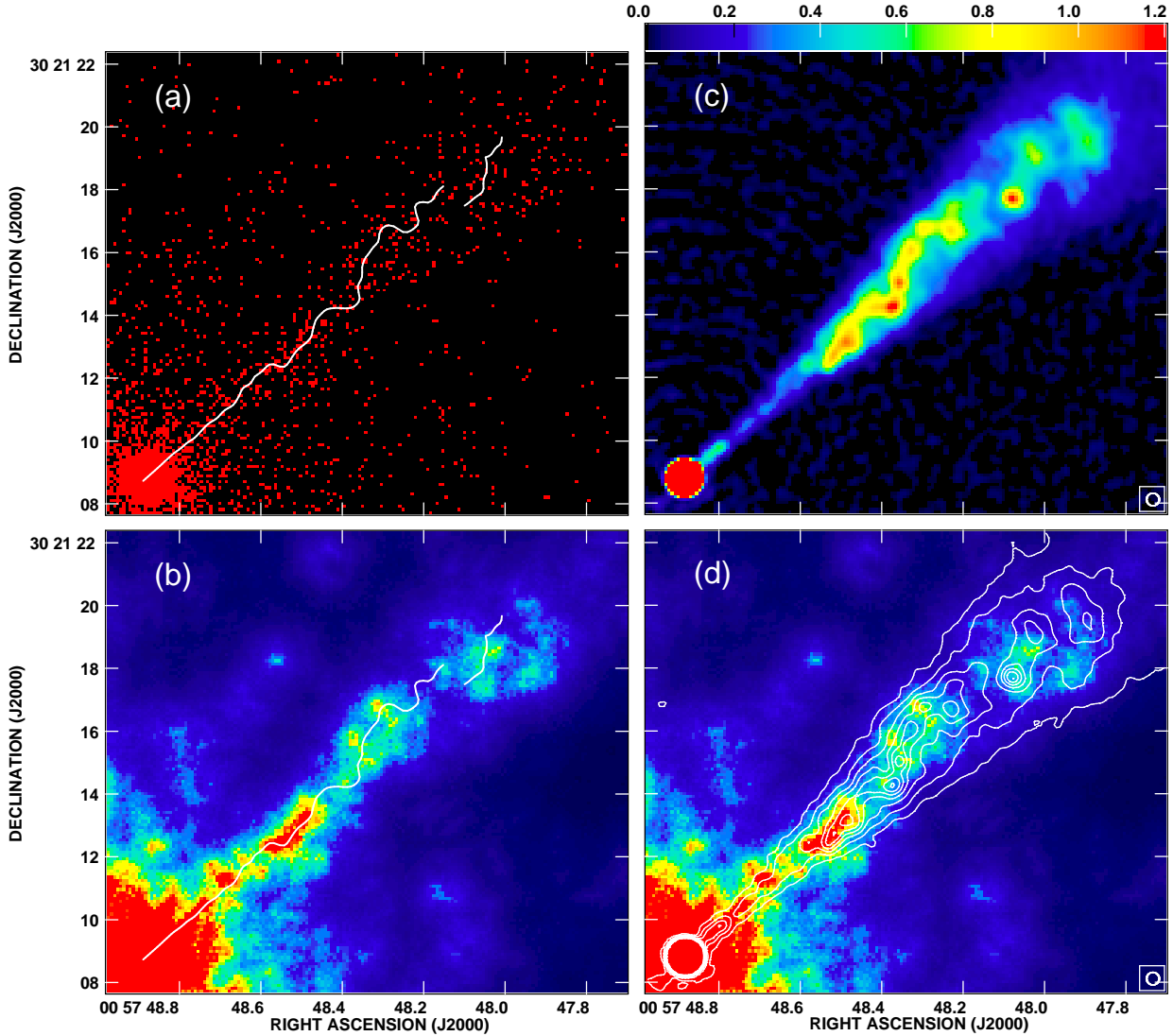}
\caption{
Plots showing the correspondence of features in the {\it Chandra\/}
0.8-5~keV X-ray data and the VLA 5~GHz data with a 0.4 arcsec beam.
(a). X-ray counts with a pixel size of 0.0984 arcsec.  The radio ridge
line, formed by connecting the maxima of Gaussian functions fitted to
a sequence of cuts taken perpendicular to the jet axis, is
superposed. (b) As (a), except that the X-ray data are adaptively
smoothed with a top-hat filter containing a minimum of 8 counts in
each smoothing kernel.  (c). Colour image of the radio data with a
scale from 0 to 1.2 mJy beam$^{-1}$.  (d). X-ray image of (b) with
radio contours at 0.05 and 0.2 mJy beam$^{-1}$, and then equally spaced
at intervals of 0.2 mJy beam$^{-1}$ up to 2 mJy beam$^{-1}$.  Note
that in (a), (b) and (d) the galaxy X-ray emission is centrally
saturated to show the X-ray jet, but see \citet{wbhn315} for an earlier radial profile of
the 0.4-7~keV galaxy emission.}
\label{fig:radioxray}
\end{figure*}

\subsection{\chandra}
\label{sec:xrayobs}

We observed \source\ in VFAINT data mode with the back-illuminated CCD
chip, S3, of the Advanced CCD Imaging Spectrometer (ACIS) on board
\chandra\ on 2003 February 22 (OBSID 4156, sequence 700835).  Details
of the instrument and its modes of operation can be found in the
\chandra\ Proposers' Observatory Guide\footnote{
http://cxc.harvard.edu/proposer}.
Results presented here use {\sc ciao
v3.2.2} and the {\sc caldb v3.1} calibration database.  We
re-calibrated and analysed the data, with random pixelization removed and
bad pixels masked, following the software ``threads''
from the \chandra\ X-ray Center (CXC)\footnote{
http://cxc.harvard.edu/ciao}. Only events with grades 0,2,3,4,6 were used.

There were some intervals during the observation when the background
rate as much as doubled, and these periods (about 5 per cent of the exposure)
were removed, leaving a calibrated dataset with an observation
duration of 51.918~ks.

The observation was made with a 512-row subarray, giving a 4 by 8
arcmin field of view in the S3 CCD. The S1, S2, and S4 CCDs were also
active. The subarray was used to reduce the readout time to 1.74~s and
so decrease the incidence of multiple events within the frame-transfer
time. In our earlier 5 ks observation \citep{wbhn315} we used a
128-row subarray with a 0.44~s readout time, and pile-up was
negligible at the core.  With the knowledge of the core count rate
from that observation, we selected the 512-row subarray for the longer
observation in order to increase the field of view while still
restricting pile-up at the core to 5 per cent.  The jet was aligned in
the 8 arcmin direction.  We took advantage of VFAINT cleaning
except for spectral analysis of the core, since piled-up events take
on the appearance of VFAINT background events and removing them would
lead to an underestimate (albeit small) of the core flux.

We shifted the X-ray image by 0.12 arcsec, mostly in right ascension,
to register the X-ray core with the radio-core position given in the
VLA Calibrator Database\footnote{
http://www.vla.nrao.edu/astro/calib/manual/csource.html},
$\alpha=00^{\rm h} 57^{\rm m} 48^{\rm s}.883$, $\delta =+30^\circ 21'
08''.81$.  This amount of shift is within \chandra's absolute aspect
uncertainties\footnote{ http://cxc.harvard.edu/cal/ASPECT/celmon/}.

\subsection{Radio}
\label{sec:radioobs}

The observations and reduction of the 5~GHz VLA radio data used
in this work are described in \citet{laing06a}.  We used intensity
images made with beams of 0.4~arcsec and 1.5~arcsec FWHM for
comparison with the X-ray emission in the inner and outer parts of the
jet, respectively.  The radio images were shifted by $\sim
0.1$~arcsec to align the core with the position given in
Section~\ref{sec:xrayobs}. We used Stokes $I$, $Q$ and $U$ images
at a resolution of 0.4 arcsec to derive the distributions of degree of
polarization and ${\bf E}$-vector position angle at 5~GHz.  The degree
of polarization was corrected to first order for Ricean bias
\citep{wardle}.
The effects of Faraday rotation were removed from
the position angles using a two-frequency rotation-measure image at
1.5~arcsec resolution \citep[fig.~11d of][]{laing06a}. Ambiguities in the
position-angle differences were determined using a five-frequency rotation-measure
fit at 5.5~arcsec resolution. The average correction for Faraday rotation over
the base of the jet is 16 degrees, and the maximum variation over the region is 2.4~
degrees. Errors in the correction for Faraday rotation are estimated to be
$<1$ degree.
Depolarization between 5 and 1.4 GHz is negligible \citep[section 5.2 of][]{laing06a},
and there is therefore no evidence either for internal Faraday
rotation or for significant unresolved gradients in foreground rotation.

\section{Jet Results}
\label{sec:results}

\subsection{Radio-jet morphology and polarization}
\label{sec:image}

The radio jet is relatively faint and unresolved in width out to about
4 arcsec from the core, after which it brightens and contains a
prominent oscillatory filament displaying a number of discrete knots
(Fig.~\ref{fig:radioxray}c).  The radio ridge line is not perfectly
straight closer to the core than 4 arcsec, but whether or not this is
an inner extension of the filament cannot be addressed by our current
data.

\begin{figure}
\centering
\includegraphics[width=3.3truein]{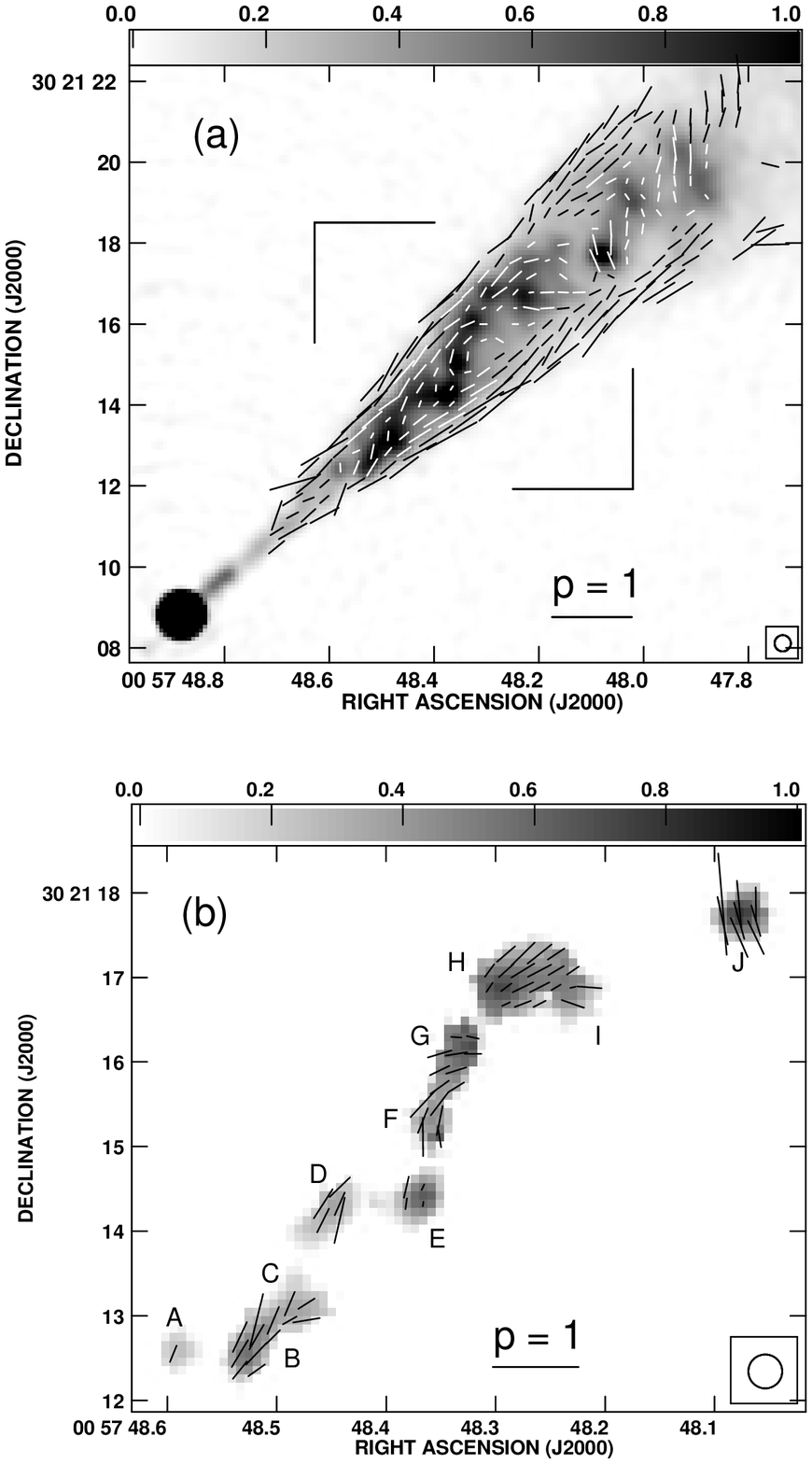}
\caption{
Vectors whose lengths are proportional to the fraction of linear
polarization, $p$ and whose directions are those of the apparent
magnetic field (i.e., perpendicular to the
E-vector position angle after correction for Faraday rotation), 
superimposed on grey-scales of total intensity at
5~GHz in the range 0 -- 1 mJy beam$^{-1}$.  The resolution is
0.4~arcsec FWHM and the intensity scales are shown by the labelled
bars. (a) An image of the inner jet covering the same area as Figure
\ref{fig:radioxray}. (b) Emission from the small-scale structure of
the filament alone, estimated as described in the
text, over the area shown by the box in panel (a).  
}
\label{fig:radiopolarization}
\end{figure}

We have examined the polarization structure of the jet and filament.
Figure~\ref{fig:radiopolarization}a shows vectors whose lengths are
proportional to the degree of polarization and whose directions are
along the apparent magnetic field, superimposed on a grey-scale of
total intensity.  \citet{canvin} concluded that the high
polarization at the edge of the jet is from a roughly equal mixture of
toroidal and longitudinal components, and that the average on-axis
field (without treating the filament as a separate entity) is close to
isotropic.

The high degree of polarization at the edges of the jet makes it
difficult to separate the polarized emission due to the filament from
that of the surrounding jet plasma. We derived a rough estimate
of the polarization of the filament alone on the assumption that the
emission from the rest of the jet is axisymmetric, as follows:
\begin{enumerate}
\item We first chose the origin of polarization position angle to be
along the jet axis so that Stokes $Q$ and $U$ (corrected for Faraday rotation
as described in Section~\ref{sec:radioobs}) are respectively symmetric and
antisymmetric under reflection in the axis for an intrinsically
axisymmetric brightness distribution.

\item We then blanked the $I$, $Q$ and $U$ images in the area covered
by the filament (estimated by eye).

\item The assumption that the remaining emission is, on average,
axisymmetric then allowed us to replace pixels in the blanked region
by their equivalents on the other side of the mid-line.
This procedure does not work for points close to the mid-line,
where we assumed instead that $Q = U = 0$ \citep[exactly true on the
mid-line for the models of][]{canvin} and interpolated $I$
from nearby pixels outside the filament area.

\item Finally, we subtracted these estimates of the diffuse emission
from the original $I$, $Q$ and $U$ data to leave images of the
filament alone.
\end{enumerate}
The results are shown in Figure~\ref{fig:radiopolarization}b, on which
the most prominent radio knots are labelled. Running along the radio
filament, the apparent magnetic-field vectors appear roughly parallel
to the filament out to component I, with knot G the main exception,
making the filament a `magnetic strand'. The typical fractional
polarization is high at $p \sim 0.2 - 0.4$ for knots A through I.
The exception is the bright knot E, where the average $p \sim 0.1$,
but there is a null in the polarization at the brightness peak,
plausibly resulting from averaging over a $90^\circ$ change in
direction of the filament if the apparent field remains aligned.
After component I, where the filament becomes poorly defined, the
vectors are roughly perpendicular to the overall jet direction, as in
the compact, radio-bright, and highly polarized component J.
There is no evidence for any anomalous Faraday
rotation or depolarization associated with the filament (see
section~\ref{sec:radioobs}).  

\subsection{The X-ray properties of the radio jet}
\label{sec:xproperties}

\begin{figure}
\centering
\includegraphics[width=3.3truein]{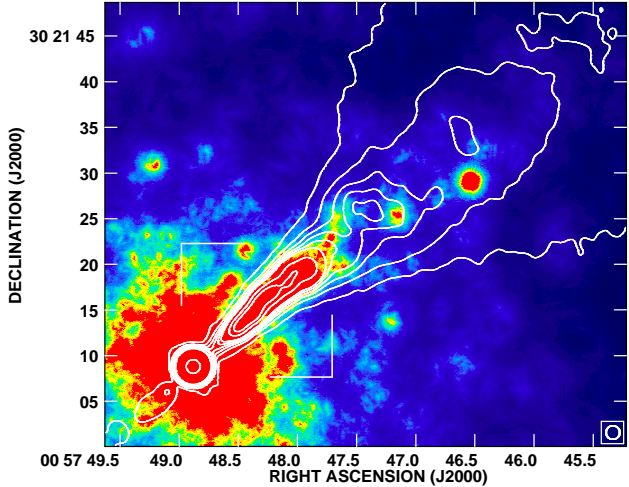}
\caption{The core and outer jet showing smoothed X-ray data as in
Figures~1b and 1d with contours of the 5-GHz radio data with
1.5-arcsec beam. The lowest contours are at 0.05, 0.4, 0.8, 1.2, 1.6
mJy beam$^{-1}$.  The white box indicates the region mapped in Figure~1.}
\label{fig:radioxrayouter}
\end{figure}

The core, galaxy atmosphere, and resolved X-ray jet that were detected
in the earlier short {\it Chandra\/} observation \citep{wbhn315}
are measured with greater precision in the new data.  Because the
galaxy atmosphere has a soft X-ray spectrum, data from the relatively
hard 0.8-5~keV X-ray band are selected for comparison with the radio
data in Figures~\ref{fig:radioxray} and \ref{fig:radioxrayouter}.

\subsubsection{X-ray and radio morphological comparison}
\label{sec:morp}

Working outwards from the core, the innermost 3 arcsec of the radio
jet cannot be matched with X-ray features because here the X-ray
emission from the core and the \source\ galaxy are relatively bright.
The first X-ray feature we identify with the jet is 3.6 arcsec from
the core at $\alpha=00^{\rm h} 57^{\rm m} 48^{\rm s}.68$, $\delta
=+30^\circ 21' 11''.3$. (Fig.~\ref{fig:radioxray}b).  To investigate
the reality of this feature and the probability that it is associated
with the jet, we extracted counts from a circle of radius 0.7
arcsec around the source.  Following \citet{wbhn315}, we 
used the {\sc iraf stsdas} task {\sc ellipse} to model the galaxy
atmosphere, based on the 0.3--2 keV data smoothed with a Gaussian of
$\sigma = 1.3$ arcsec, and used the resulting intensity contours
to define an elliptical annulus of full width 1.4 arcsec, centred on
the core, in which the source circle falls.  At 0.8--5 keV, the source
circle contains 40 counts, whereas the average value for a circle of
the same size lying anywhere in the annulus is 14.7 counts, making the
detection highly significant.  The largest number of counts in a
circle of the same size positioned anywhere else in the annulus is 23,
and therefore it is unlikely to be a coincidence that the circle with 40
counts lies in projection on the radio jet.  This inner X-ray knot
lies in a region of weak extended radio emission from the inner jet,
and there appears to be no significantly enhanced radio emission at
its position.

The complex incorporating knots A through D in
Figure~\ref{fig:radiopolarization}b is bright in both X-ray and radio
(Fig.~\ref{fig:radioxray}d), and shows particularly close structural
similarities.  Throughout the region to knot I there is distributed
X-ray emission over the full extent of the diffuse radio jet, and
localized X-ray peaks follow the ridge line.

The outermost structure in Figure~\ref{fig:radioxray}, incorporating
knot J in Figure~\ref{fig:radiopolarization}b), is where the radio
filament shows considerable brightness fluctuations, and by this point
the coherent structure has been lost, with the filament broken into
pieces.  The filament appears to be breaking up where the jet broadens
in both radio and X-ray, and the flow is thought to slow down.  The
general correspondence between the radio and X-ray emission is still
good, despite the poorer X-ray photon statistics.  However, the
bright, compact radio component J ($\alpha = 00^{\rm h} 57^{\rm m}
48^{\rm s}.1$) has only a faint X-ray equivalent, despite being as
bright as knot E at 5 GHz.

In the outer reaches of the X-ray jet (Fig.~\ref{fig:radioxrayouter})
where coherent and compact radio features cease, we nonetheless see a
correspondence between brighter X-ray and radio emission as far out as
the feature at $\alpha = 00^{\rm h} 57^{\rm m} 47^{\rm s}.1$.  The
X-ray source at $\alpha =00^{\rm h} 57^{\rm m} 46^{\rm s}.5$ has no
particular relationship with the radio jet and is likely to be a
background object.  However, the faint structure at $\alpha =00^{\rm
h} 57^{\rm m} 47^{\rm s}.1$ and the diffuse emission around it (about
$30''$ from the nucleus) do appear to be jet related, and mark the
furthest extent to which we are confident that X-ray jet emission is
detected.  Thus although the X-ray emission is brightest in the first
14 arcsec of the jet, it is also detected in the region between 14 and 30
arcsec where the jet is decelerating.

\subsubsection{Jet spectrum and Intensity}
\label{sec:jetxpec}

\begin{figure}
\centering
\includegraphics[width=3.3truein]{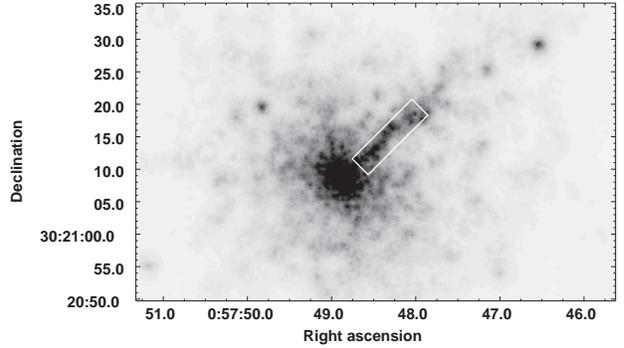}
\caption{
Image showing 0.3-2 keV counts in 0.0984 arcsec pixels, adaptively
smoothed with a top-hat filter containing a minimum of 8 counts in
each smoothing kernel.  The on-source region used to measure the
spectrum of the jet is shown.  The softer X-rays shown here, as
compared with Figures~\ref{fig:radioxray} and
\ref{fig:radioxrayouter}, emphasize emission from the galaxy gas.
} 
\label{fig:jetregion}
\end{figure}

\begin{figure}
\centering
\includegraphics[width=3.3truein]{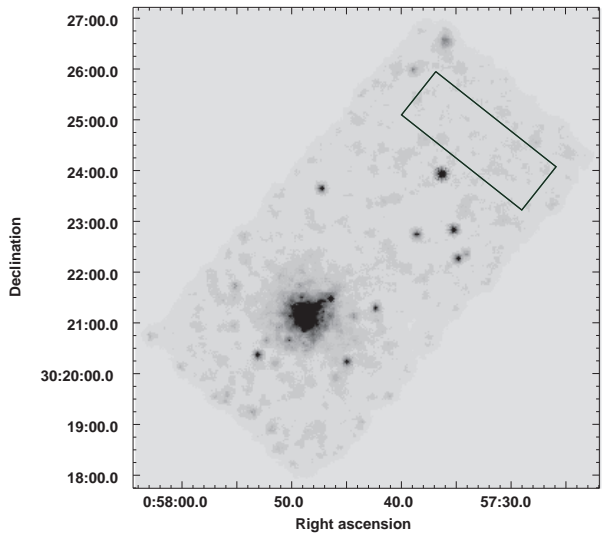}
\caption{
The exposed part of the ACIS-S3 chip.  The 0.3-7~keV X-ray counts in
1.476 arcsec pixels have been exposure corrected and adaptively
smoothed with a top-hat filter containing a minimum of 15 counts in
each smoothing kernel.  The jet and gas, and unassociated point
sources are seen.  The rectangle marks the region used to measure the
background for spectral fitting of the gas.  Unresolved X-ray emission
is seen from the flat-spectrum radio source at $\alpha=00^{\rm h}
57^{\rm m} 38^{\rm s}.720$, $\delta =+30^\circ 22' 44''.99$ \citep{laing06a},
confirming the earlier suggestion of \citet{wbn315}
based on \rosat\ and Palomar Sky Survey data that,
although the source lies in projection on the large-scale radio jet,
it is an unrelated background object.
} 
\label{fig:frame}
\end{figure}

We used the on-source region shown in Figure~\ref{fig:jetregion} and
the background region shown in Figure~\ref{fig:frame} to measure the
composite spectrum of the X-ray jet in the region containing the radio
filament, between 3.2 and 16.2 arcsec from the nucleus.  There are 696
net counts between 0.3 and 5~keV.  The emission has a soft spectrum.
A fit to a single-component absorbed power law is poor ($\chi^2 = 34$
for 18 degrees of freedom).  As anticipated, since the jet is
projected on significant thermal X-ray emission, an improvement was
obtained when a thermal model was included, accounting for about 22
per cent of the counts. A fit with the only source of absorption being
that in our own Galaxy ($N_{\rm H} = 5.92 \times 10^{20}$ cm$^{-2}$;
\citet{dickey} gives $\chi^2 = 17.4$ for 15 degrees of
freedom.

\begin{figure}
\centering
\includegraphics[width=2.5truein]{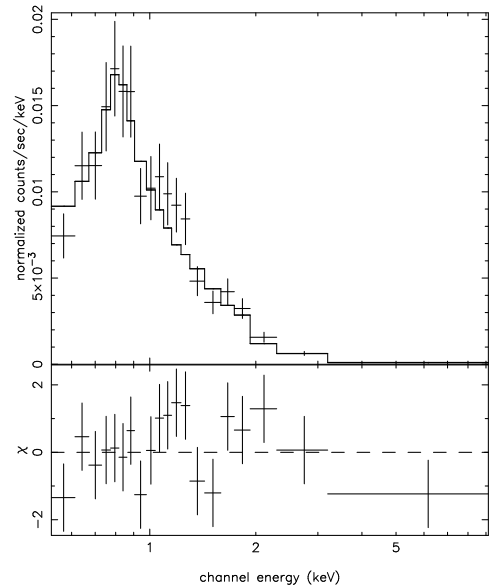}
\caption{
The X-ray spectrum of composite emission from the jet fitted to
the combination of a power-law and thermal model.
} 
\label{fig:jetspectrum}
\end{figure}

\begin{figure}
\centering
\includegraphics[width=1.7truein]{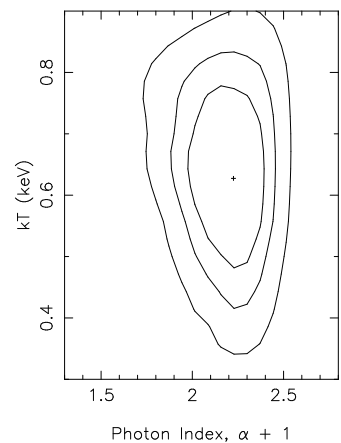}
\caption{
$\chi^2$ contours (1$\sigma$, 90 per cent and 99 per cent, for two
interesting parameters) showing the fitted uncertainties in
power-law spectral index and $kT$ for counts
extracted from the jet region.  The thermal abundances have been
constrained to lie between 10 and 100 per cent of solar, and
absorption is fixed to the Galactic value.
} 
\label{fig:jetspeccontour}
\end{figure}

The jet X-ray spectrum is shown in Figure~\ref{fig:jetspectrum}.  The
best-fit gas abundance is 0.25 solar, but while this parameter is very
poorly constrained the other parameters of the fit (except for the gas
emission measure) are insensitive to the value adopted.  Uncertainties
in $kT$ and power-law index (Fig.~\ref{fig:jetspeccontour}) are found
with the abundance constrained to lie between 10 and 100 per cent of
solar.  The gas temperature is consistent with that found generally
for the \source\ galaxy (see Sec.~\ref{sec:gasxpec}).  The jet
spectrum, with its best-fit slope of $\alpha = 1.2$ is relatively
steep, and steeper than the core spectrum (see
Sec.~\ref{sec:corexpec}).  $\chi^2$ decreases slightly if a small
level of intrinsic absorption, as may occur from cold gas in the
\source\ galaxy, is included in the fits, and then the power-law index
increases by a few tenths.  While the statistics are insufficient to
claim that this excess absorption is significant, we can conclude at
$\sim$~3$\sigma$ confidence that the power-law index is no flatter
than $\alpha = 0.9$.

The radio spectral index in the region over which the X-ray spectrum
has been extracted is $\alpha_{\rm r} = 0.61$,
whereas our best estimate for $\alpha_x$ is $1.2\pm 0.2$.  The X-ray
luminosity in the region in which the jet spectrum has been extracted
is $(4.3 \pm 0.2) \times 10^{40}$ ergs s$^{-1}$ (0.3-5 keV), and the
1~keV flux density corrected for Galactic absorption is
$10.5^{+2.0}_{-2.9}$~nJy.  This is more accurately determined than our
previously published value because we now have an X-ray spectral
measurement.  The corresponding radio flux density of 74 mJy leads to
a radio-to-X-ray spectral index of $\alpha_{rx} =
0.89$.  While the jet has yet to be detected in the optical, $\alpha_x
> \alpha_{rx} > \alpha_r$ requires, in the simplest case, a broken
power-law spectrum, and supports a synchrotron origin for the X-ray
emission.  Although the optical upper limits from the HST imaging of \citet{verdoes}
lie significantly above an extrapolation of
the radio spectrum between 1.4 and 5 GHz, the upper limit of $3.2
\mu$Jy at 4500\AA\ of \citet{butcher}, based on ground-based observations,
requires the spectrum to steepen between the radio and optical.

The radio flux density of the filament alone (estimated from the image
shown in Fig.~\ref{fig:radiopolarization}b, convolved to a resolution
of 0.6~arcsec FWHM to match the \chandra\ point-spread function) is
7.3~mJy, which is 10 per cent of the radio flux from the rectangular
region shown in Figure~\ref{fig:jetregion} that contains 696 net X-ray
counts (0.3--5~keV).  The convolved radio image was used to define a mask covering
the area of the filament, and 184 X-ray counts (0.3-5 keV) were summed
over the mask.  Since the mask allows through both filament and
diffuse-jet X-ray counts, it was necessary to estimate and subtract
the background contributed by the diffuse jet by sampling regions
within the jet envelope that are adjacent to the filament.  The wiggly
nature of the filament means we sampled diffuse regions at similar
transverse distances from the jet axis as the filament.  The
background estimate was 94 counts, giving a statistically significant
90 net X-ray counts.  Within the rather large statistical
uncertainties (13 per cent), and harder to quantify systematic
uncertainties in the diffuse-jet contribution, this is consistent with
being the same fraction of total jet emission as in the radio.

\begin{figure}
\centering
\includegraphics[width=2.7truein]{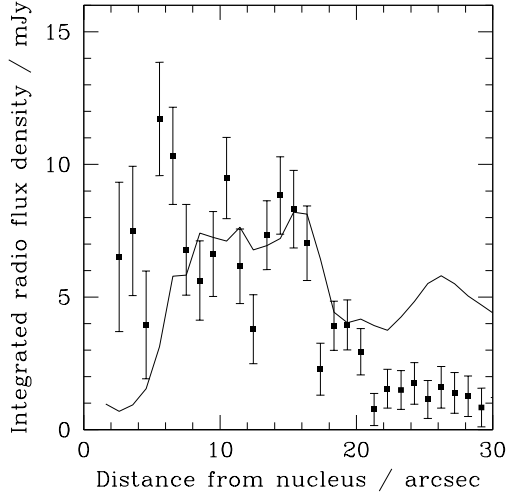}
\caption{
Profiles along the jet of the 5-GHz radio flux density (solid line)
and 0.3-5 keV X-ray net counts (data points with Poisson errors and
arbitrary scale).  The radio profile is from a map with a restoring
beam of 0.6 arcsec FWHM, to match the \chandra\ resolution.  The width
of the on-source extraction boxes are set interactively using the size
of the jet envelope in the radio image.  The X-ray background
measurements are from regions of $\sim5$-arcsec width on either side
of the jet.  Additional uncertainties apply to the inner few X-ray
bins due to imperfect subtraction of the galaxy counts.  }
\label{fig:xrprofile}
\end{figure}

\begin{figure}
\centering
\includegraphics[width=2.4truein]{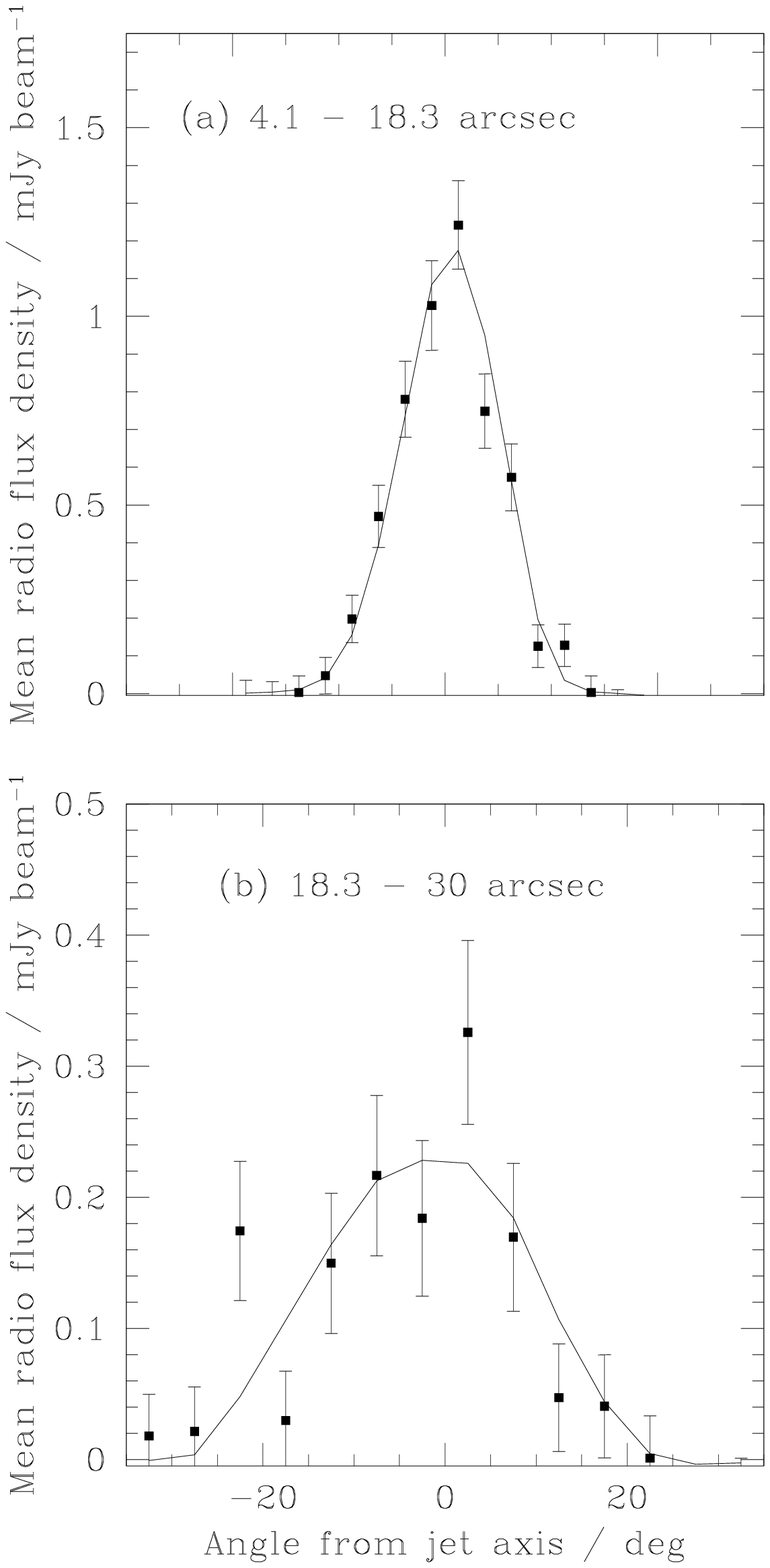}
\caption{
As in Figure~\ref{fig:xrprofile} but across the jet.  The integrations
along the jet are for distances from the nucleus of (a) 4.1 to 18.3
arcsec, where the filament contributes about 10 per cent of the flux
and (b) 18.3 to 30 arcsec, where the jet is believed to be
decelerating, and where the X-ray to radio flux ratio has decreased
significantly. The arbitrary scale for the X-ray points is the same as
in Figure~\ref{fig:xrprofile} for panel (a), and lower by a
factor of 2.7 in (b).
 }
\label{fig:xrtranprofile}
\end{figure}

A comparison of the profiles of X-ray and radio emission down the jet
are shown in Figure~\ref{fig:xrprofile}.  The X-ray emission drops
relative to the radio beyond 16 arcsec where the jet is decelerating.
There is reasonable agreement in jet width between the radio and X-ray
in both the fast and decelerating regions of the jet
(Fig.~\ref{fig:xrtranprofile}), although the X-ray statistics in the
decelerating region are relatively poor.  An implication is that the
processes which govern the relative amounts of diffuse X-ray and radio
emission depend primarily on distance from the nucleus and do not vary
significantly across the jet.

\section{Other X-ray components}
\label{sec:nonjetresults}

\subsection{The central region}
\label{sec:corexpec}

We extracted the X-ray spectrum for a circle of radius 1 arcsec
centred on the core, using the background region shown in
Figure~\ref{fig:frame}.  In contrast to the situation for the short
observation, where a single-component power law gave a good fit, the
improved statistics of the new data require the inclusion of a thermal
component, as found also by \citet*{donato}.  6.6 per cent of the
counts (significant only below 1 keV) arise from the thermal component
in our best composite fit, for which $\chi^2 = 103$ for 94 degrees of
freedom.  Although the abundance of the gas is poorly constrained, the
best-fit value is solar, and this was frozen in the estimates of
the uncertainties on the other parameters in this fit.

\begin{figure}
\centering
\includegraphics[width=2.5truein]{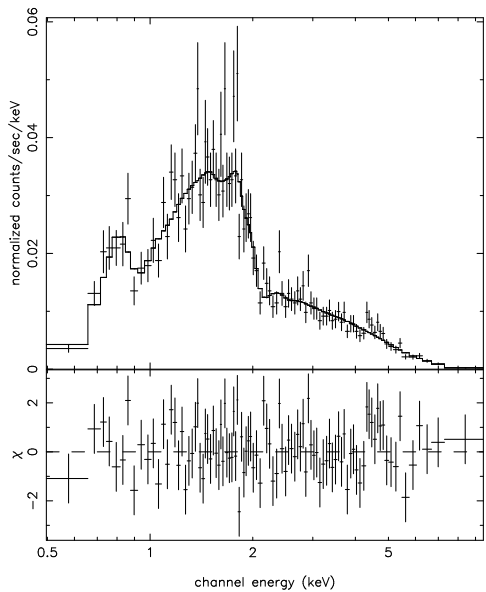}
\caption{
The X-ray spectrum of composite emission from the core fitted to the
combination of a power-law and thermal model.  }
\label{fig:corespectrum}
\end{figure}

\begin{figure}
\centering
\includegraphics[width=1.6truein]{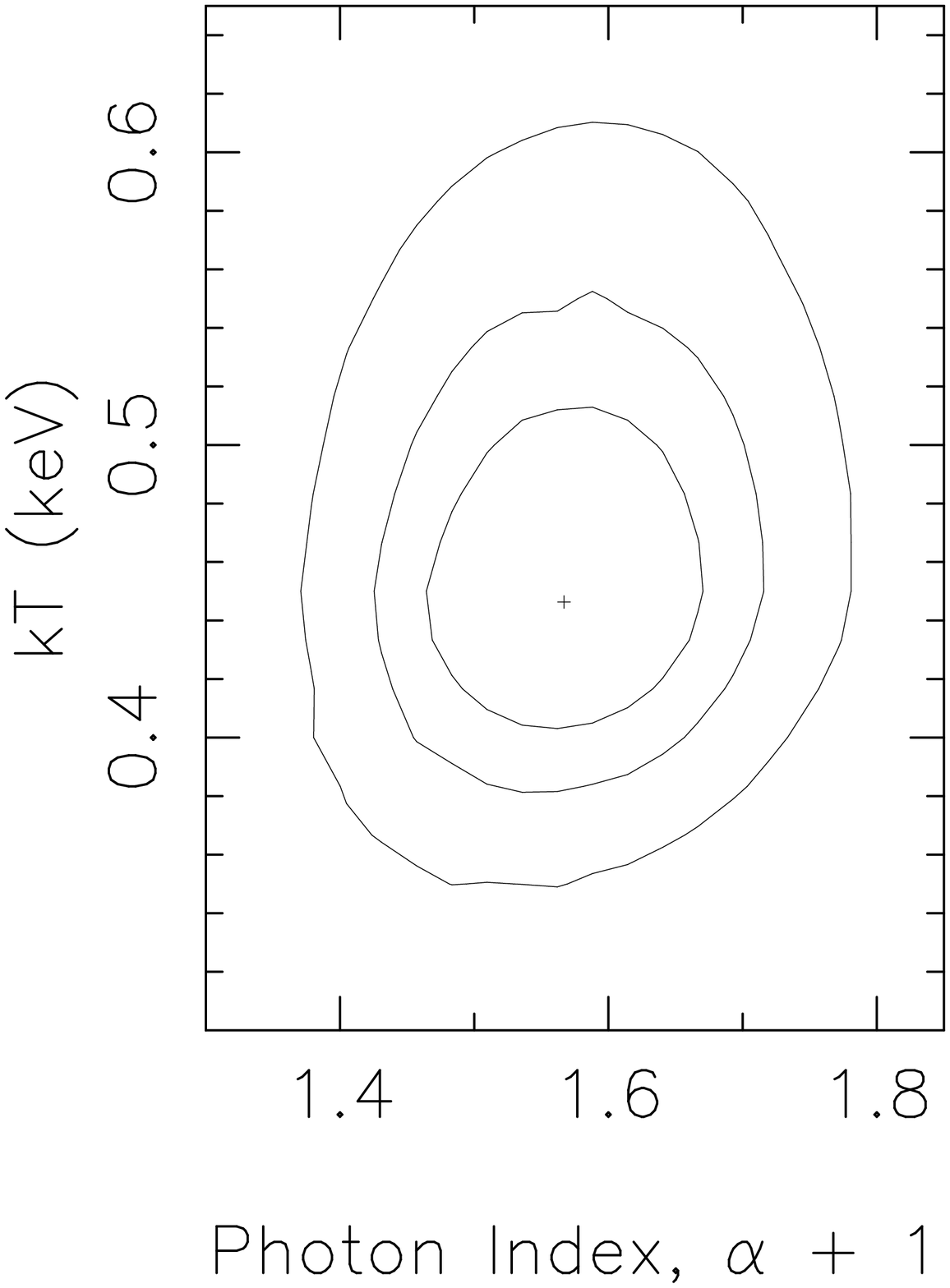}
\includegraphics[width=1.6truein]{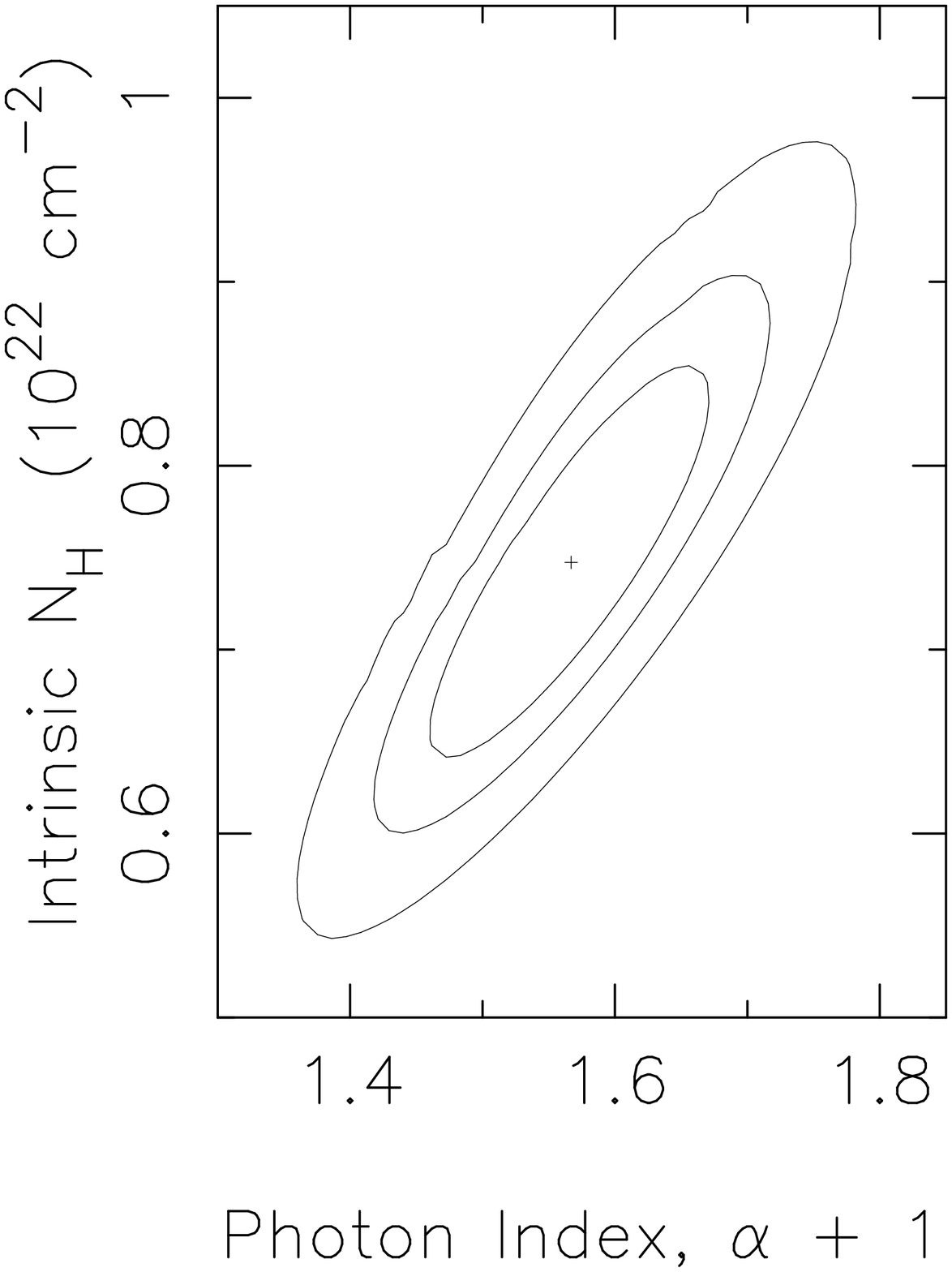}
\caption{
$\chi^2$ contours (1$\sigma$, 90 per cent and 99 per cent, for two
interesting parameters) showing uncertainties in parameters in the
two-component fit to the X-ray data from a 1-arcsec-radius region
centred on the core.  }
\label{fig:corecont}
\end{figure}

The X-ray spectrum of the core (Fig.~\ref{fig:corespectrum}) is
strikingly harder than that of the jet (Fig.~\ref{fig:jetspectrum}),
and the fitted power-law index is indeed significantly flatter than
that of the jet, taking into account the uncertainties (compare left
panel of Fig.~\ref{fig:corecont} with Fig.~\ref{fig:jetspeccontour}).
Application of the pileup model available in {\sc xspec\/} finds that
pileup (estimated to be $\sim 5$ per cent) has negligible effect on
our spectral fits.  Table~\ref{tab:powerlaws} gives the fitted
power-law model parameters for both the core and jet regions.

\begin{table}
\caption{Parameter values for the power-law spectral fits}
\label{tab:powerlaws}
\begin{tabular}{lcll}
Structure & $\alpha_x$ & $n_{\rm H}$ (cm$^{-2}$)$^{\rm a}$ & $f_{\rm 1~keV}$ (nJy) \\
Core & $0.57 \pm 0.11$ &$(7.6 \pm 1.2) \times 10^{21}$& $120 \pm 20$\\
Jet$^{\rm b}$ &$1.2\pm 0.2$ &0~f& $10.5^{+2.0}_{-2.9}$\\
\end{tabular}
\medskip
\begin{minipage}{\linewidth}
90\% uncertainties for 1 interesting parameter ($\chi^2_{\rm min} +
2.7$). For both structures a thermal component is also included in the
spectral fits.  a.  Excess over Galactic column density. f = fixed.
b. For the region shown in Figure~\ref{fig:jetregion}.
\end{minipage}
\end{table}

In agreement with our earlier results and those of \citet{donato}, 
intrinsic absorption of $\sim 7.6 \times 10^{21}$
cm$^{-2}$ is required for the power-law component
(Fig.~\ref{fig:corecont}) which, corrected for absorption and the
point spread function, has a 1 keV flux density of $0.12 \pm 0.02
~\mu$Jy (90\% uncertainty for 1 interesting parameter) and a 0.5-8~keV
luminosity of $(7.6\pm 0.3) \times 10^{41}$ erg s$^{-1}$.  The
relatively low intrinsic absorption supports our earlier conclusions
that the measured core emission from this source is predominantly
jet-related in origin \citep{wb, wbhn315}.
\source\ is a dusty galaxy with a resolved disk-like
structure (measured with the Hubble Space Telescope) that is responsible for a
reddening of $A_{\rm v} = 0.25$ \citep{deruiter}.  We can use
the expression given in \citet{wilkes} \citep[based on][]{burstein}
to convert reddening into hydrogen column density under the
assumption of a Galactic dust-to-gas ratio, and find $N_{\rm H} = 7
\times 10^{20}$ cm$^{-2}$, insufficient by an order of magnitude to
account for the X-ray absorption.  However, \citet{deruiter} also
estimate a dust mass from IRAS 100~$\mu$m data that is a thousand
times larger than that inferred from the reddening absorption study,
and \citet{leon} report the detection of carbon monoxide with a
double-horned line profile characteristic of a rotating disk or torus,
and estimate a molecular gas mass of $\sim 3 \times 10^8$ M$_\odot$.
The precise nature and location of most of the gas and dust in
\source\ is currently unknown, and some could be associated with an
inner torus. While (weak) broad emission lines have been detected
\citep{ho}, the results do not constrain the properties of a
possible torus, as lines are seen in both direct and
reflected (polarized) light \citep*{barth}.

If we wish to measure the accretion efficiency of the central source
we should allow for the possible presence of a torus to prevent
underestimating the central luminosity.  Even with a fitted column
density of less than $10^{22}$ cm$^{-2}$, the X-ray observations do not
rule out the presence of an inner torus with a line-of-sight column
density of, for example, $10^{23}$ cm$^{-2}$; they merely set an upper
limit on the X-ray luminosity that can lie behind such a putative
torus \citep[e.g.,][]{wb346, evans}.  The high
quality of the X-ray dataset for \source\ allows us to determine
whether this luminosity is consistent with that expected for accretion
at close to the Eddington limit onto a standard, geometrically thin
and optically thick accretion disk (Shakura-Sunyaev or similar) or
whether a radiatively inefficient accretion flow
\citep[e.g.,][]{narayan, blandford} is suggested.  In the case of
\source, for a photon spectral index of 2.0 we find that any concealed
X-ray component has 90\%-confidence upper limits at 0.5-8 keV of $7
\times 10^{41}$ or $1.5 \times 10^{42}$ erg s$^{-1}$ if absorbed by
gas with $N_{\rm H} = 10^{23}$ or $10^{24}$ cm$^{-2}$,
respectively. Since empirical relations find a black-hole mass of
$\sim 2 \times 10^9$ M$_\odot$ \citep{bettoni,marchesini}, the limits
must be compared with an Eddington luminosity of
$\sim 2.6 \times 10^{47}$ erg s$^{-1}$. Even allowing for output
at wavelengths other than the X-ray, accretion in \source\ must be
greatly sub-Eddington or radiatively inefficient.

Scaling the 3690 counts (0.4-7 keV) from the core by observing time,
we would have expected $329\pm 5$ counts in our earlier short
observation, in agreement with the $313\pm 18$ counts detected
\citep{wbhn315}.  Thus we have no indication that the core has
varied in intensity over the 28 months between the
observations. However, \citet{fab92} and
\citet{wbn315} present evidence that the X-ray core
intensity has varied historically, and the radio flux density from the
parsec-scale core has shown considerable variation in the recent past
\citep{cotton99, lazio}.

The central gas as measured from the core spectrum
(Fig.~\ref{fig:corecont}) is relatively cool as compared with the rest
of the galaxy gas, consistent with a cooling time that is short
compared with the Hubble time (Sec.~\ref{sec:gasxpec}).  We find a
density of 0.28 cm$^{-3}$ and pressure\footnote{1 Pa = 10 dyne
cm$^{-2}$} of $4.5 \times 10^{-11}$~Pa when we model the gas as a
uniform sphere of radius 1 arcsec.  The consistency of this pressure
with that of adjacent regions where the contribution from the core is
small (Sec.~\ref{sec:gasxpec}) lends supporting evidence that we are
applying the correct composite model to the core emission.

\subsection{Properties of the X-ray-emitting gas}
\label{sec:gasxpec}

The deeper observation allows the temperature and distribution of the
diffuse X-ray-emitting gas around \source\ to be measured with greater
precision.  As in \citet{wbhn315} we extracted a radial
profile and fitted it to the combination of a point source (for the
core) and $\beta$-model\footnote{Surface brightness proportional to
$[1 + (\theta^2/\theta_{\rm cx}^2)]^{0.5 - 3\beta}$, where
$\theta_{\rm cx}$ is the core radius \citep{cav, sarazin}}, both convolved with the point spread
function. Out to a radius of about 40 arcsec the gas is well described
by the $\beta$ model, but beyond this it drops off more sharply.  This
drop-off will be discussed elsewhere, together with the implications
for the dynamics of the jet.

\begin{figure}
\centering
\includegraphics[width=1.7truein]{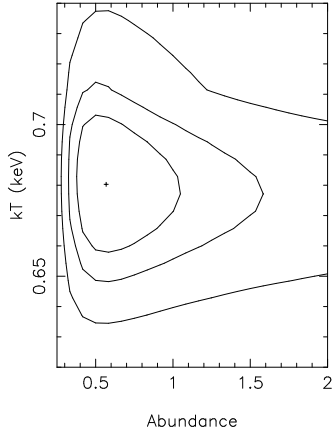}
\caption{
$\chi^2$ contours (1$\sigma$, 90 per cent and 99 per cent, for two
interesting parameters) showing uncertainties in parameters in the
spectrum of the X-ray gas from a core-centred annulus of radii 4 and
20 arcsec and excluding a 30-degree pie slice around the
jet. $\chi^2_{\rm min} = 41$ for 47 degrees of freedom.  }
\label{fig:gascontour}
\end{figure}

Out to the maximum radius of the resolved X-ray jet the gas is fitted
with a $\beta$ model of core radius $1.7 \pm 0.2$~arcsec and $\beta =
0.52 \pm 0.01$ (1 $\sigma$ uncertainties for two interesting
parameters).  We can convert the distribution of integrated counts to
an intrinsic profile of pressure using equations given in
\citet{birkinshaw} \citep[for further details see][]{wbrev} if the gas
is isothermal.  However, here there is a complication.  The counts
extracted from a source-centred annulus of radii 42 and 120 arcsec,
using the region in Figure~\ref{fig:frame} as background, give a good
fit ($\chi^2 = 78$ for 77 degrees of freedom) to a thermal model with
$kT = 0.86_{-0.06}^{+0.14}$ keV (90\% uncertainties for 1 interesting
parameter) which is significantly hotter than in the central region
(Fig.~\ref{fig:corecont}).  To estimate the temperature of gas at
radii where the resolved X-ray jet is best measured, we used an
on-source annulus of radii 4 and 20 arcsec (appropriate for projected
distances along the jet between 2.5 and 12 arcsec for an adopted angle
to the line of sight of 38 degrees) with background measured from an
adjacent annulus of radii 20 and 42 arcsec, in each case excluding a
pie slice of 30 degrees around the jet direction.  The data fit mostly
a thermal model, but include a small contribution from the wings of
the point spread function that is fitted with a power law of spectral
index and absorption consistent with those found from the central
1-arcsec circle. Uncertainties in the temperature and abundance are
shown in Figure~\ref{fig:gascontour}.  The gas here is at a
temperature between that in the core and that in the outer reaches of
the gas distribution.  Table~\ref{tab:gastemps} is a summary of the
measured temperatures for different angular radii from the nucleus.

\begin{table}
\caption{Gas temperatures}
\label{tab:gastemps}
\begin{tabular}{lcll}
Region & Radii from nucleus & $kT$ (keV)$^{\rm a}$ & $Z_\odot^{\rm a, b}$\\
& (arcsec) \\
Core$^{\rm c}$ & $< 1$ & $0.44^{+0.08}_{-0.04}$ & 1~f\\
Inner gas$^{\rm d}$ & 4 -- 20 &$0.68 \pm 0.03$& $0.6^{+0.55}_{-0.25}$\\
outer gas & 42 -- 120 & $0.86^{+0.14}_{-0.06}$& $0.25^{+0.47}_{-0.14}$ \\
\end{tabular}
\medskip
\begin{minipage}{\linewidth}
a. 90\% uncertainties for 1 interesting parameter ($\chi^2_{\rm min} +
2.7$). b. Abundances as a fraction of Solar. f = fixed.
c. Model-fit to data from region includes power-law core emission.
d. Region excludes that of the X-ray jet.
Model-fit to data includes small component of power-law emission from the
wings of the point spread function describing the nucleus.
\end{minipage}
\end{table}

\begin{figure}
\centering
\includegraphics[width=3.0truein]{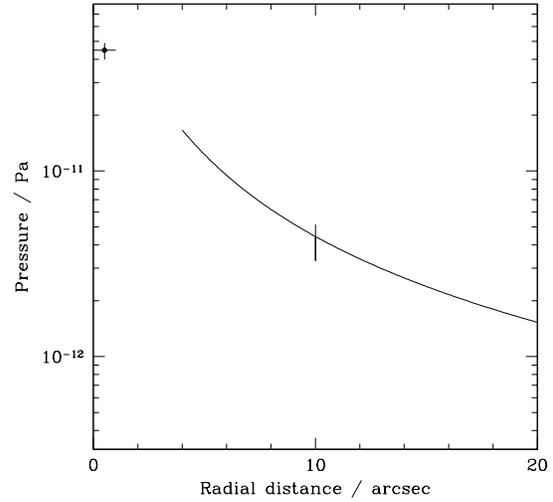}
\caption{ Pressure of the X-ray emitting gas.  The point at 0 to 1
arcsec radius is determined from the spectral fit to the core,
assuming the gas component is from an isotropic sphere.  The rest of
the profile combines our beta-model fits with spectral parameters from
Figure~\ref{fig:gascontour}.  The representative $1\sigma$ error bar
takes into account uncertainties in spectral and spatial
parameters.  }
\label{fig:gaspressure}
\end{figure}

While there is a significant degree of non-isothermality in the gas
distribution, the pressure structure is dominated by the change in gas
density with radius, and the assumption of a constant temperature
is an adequate description of
the jet's environment between 4 and 20~arcsec.  
Figure~\ref{fig:gaspressure} shows the run of gas pressure with
distance from the core. 
The external gas pressure will be combined
with the kinematical models of \citet{canvin} into a dynamical
model for the jet in a forthcoming paper.
The gas cooling time is short, at $2.8
\times 10^9$ years and $4 \times 10^8$ years at 15 arcsec ($\sim
5$~kpc) and 4 arcsec ($\sim 1.3$~kpc) from the core, respectively.

\section{Discussion}
\label{sec:discussion}

\subsection{Particle-acceleration constraints from the X-ray emission}
\label{sec:acceleration}

The radio jet emission is well established as being synchrotron
radiation from a plasma whose relativistic bulk motion produces the
initial brightness asymmetry between the jet and the counterjet.  The
physical coincidence and morphological similarity of the X-ray and
radio jets strongly suggest that they share a common origin within the
relativistic bulk flow.  Inverse-Compton models applied to the
radio-emitting electron population for the dominant sources of
scattered photons (jet synchrotron, cosmic microwave background),
assuming minimum energy, significantly underpredict the observed X-ray
emission.  
For the diffuse jet region over which the X-ray spectrum is
extracted, the magnetic field strength, $B$, would need to be at
least a factor of 45 below the minimum-energy value, $B_{\rm me}$,
increasing the total energy in the source by at least a factor
of 280 as compared with minimum energy.
This is in contrast to the range  $0.3 B_{\rm me} < B < 1.3 B_{\rm me}$
typically measured for diffuse radio-emitting plasma
where the X-ray emission {\it is} reliably associated with inverse
Compton scattering \citep{croston}.
The results therefore support a synchrotron origin for the X-ray
emission, which thus marks regions where large numbers of electrons
can be accelerated to high energies.

The X-ray knot closest to the core, at a distance of 4 arcsec, lies
where the radio emission is still relatively weak but slowly
increasing, before the abrupt brightening.  \source\ is therefore
similar to 3C~66B \citep*{hard01}, 3C~31 \citep{hard02} and 3C~ 15
\citep{kataoka03,dulwich} in showing X-ray emission in the radio-faint
region immediately before the abrupt radio brightening.  In the
symmetrical, relativistic model of \source\ developed by
\citet{canvin}, the increase in rest-frame emissivity which leads to
the radio brightening occurs at the same distance from the nucleus in
the main- and counter-jet. This suggests that the brightening is
related to a larger-scale phenomenon, perhaps associated with changes
in external pressure. The occurrence of significant X-ray emission
before the radio brightening complicates the simple picture.

There is also a clear variation in the ratio of X-ray to radio
emission along the main jet at larger distances
(Fig.~\ref{fig:xrprofile}). The ratio decreases by a factor $\sim 3$
at a projected distance of about 16~arcsec from the nucleus,
consistent with the start of deceleration \citep[$\sim 14$~arcsec in
the model of][]{canvin}. X-ray emission is detected throughout the
rapid deceleration zone (14 -- 33~arcsec) but not beyond.  3C\,31
shows a similar behaviour \citep{laing04}, and a decline in X-rays is
seen along the jet of 3C\,66B \citep{hard01} although here no
deceleration models have been constructed. Thus there appears to be a
link between the speed of the jet flow and the X-ray/radio ratio, and
consequently to the jet's particle acceleration.  
Wherever X-ray emission is detected in 3C~31 and \source, the
variation of radio
emissivity with distance from the nucleus is inconsistent with models
in which
the relativistic particles change energy only by adiabatic losses and
the magnetic field is frozen into the flow \citep{laing04,canvin}.
This provides independent evidence that dissipative effects are
important, consistent with the requirement for particle acceleration.

The occurrence of X-ray emission throughout the jet volume implies
that there must be distributed particle acceleration. The lifetime
problem is particularly acute in the filament, where the X-ray
emission is brightest: for knot E's minimum-energy magnetic field of
$\sim$4~nT, the lifetime for the X-ray-emitting electrons is $\sim$400
yr. This is comparable with the light-travel time across the knot, but
far smaller than the equivalent for the entire X-ray emitting region
($\sim$50,000 yr).  As the flow in FRI jets is probably internally
transonic \citep{bick2,laing02b}, Fermi acceleration at strong shocks
is not a plausible acceleration mechanism. It is more likely that
strong shear associated with jet deceleration causes a small fraction
of the kinetic energy of the jet to be fed, by turbulence in the shear
layer, into the high-energy electrons responsible for the X-radiation
\citep[e.g., as in][]{stawarz}.

The synchrotron minimum-energy pressures of individual knots may
slightly exceed the pressure of the external X-ray-emitting medium.
For example, knot E's internal minimum-energy pressure is $\sim 4
\times 10^{-12}$~Pa if the knot is moving with 0.9$c$ at 38 degrees to
the line of sight (higher if moving more slowly) as compared with the
external pressure of $\sim 3 \times 10^{-12}$~Pa at its deprojected
radius of 14~arcsec. A preliminary conservation-law analysis (work in
preparation) finds that the diffuse radio emission at the start of
rapid expansion in the jets is also overpressured, as found earlier
for 3C\,31 \citep{laing02b}.

\subsection{The nature of the filament}
\label{sec:filament-nature}

Complex, non-axisymmetric structure appears to be common in the
jet-brightening (and geometrically flaring) regions of FRI jets
observed with
sufficiently high angular resolution. In addition to \source, good
examples are 3C\,31,
Cen~A, B2\,0326+39 and  3C\,296
\citep{laing02a,hard03,cl04,laing06b}.  This 
non-axisymmetric radio emission is seen: 
\begin{enumerate}

\item downstream from where enhanced X-ray emission associated with
the jet is detected, and thus where there is the first evidence of
local particle acceleration, and 

\item before the region where significant jet deceleration begins, in
sources where this has been modelled.

\end{enumerate}

There is no evidence for variation in the average X-ray/radio ratio
across the jet in \source\ (Fig.~\ref{fig:xrtranprofile}), although
the constraints are poor close to the nucleus (because of inadequate
resolution) and beyond $\ga 18$~arcsec where the X-ray emission is
faint.  Neither does the average X-ray/radio ratio differ grossly
between the filament and the surrounding diffuse emission
(Sec.~\ref{sec:jetxpec}), despite differences in fine-scale structure
between the two wavelength bands. It therefore seems most likely that
the filament is an emissivity enhancement within a region of
distributed particle acceleration rather than a favoured location for
particle acceleration.

The filament in \source\ differs both morphologically and in its
polarization properties from features which have been identified as
strong shocks, such as those producing bright X-rays in M~87
\citep{perlman} and 3C~15 \citep{dulwich}. The latter tend to show
apparent magnetic fields transverse to the jet axis, whereas the
filament has a primarily longitudinal field. Although the {\it
non-axisymmetric\/} structures observed in FRI sources other than
\source\ are not well enough measured or resolved to separate their
apparent field structures from those of the surrounding emission, as
we have done for \source, in several sources the composite
polarization data are consistent with apparent fields aligned with the
structures.

\begin{figure}
\centering
\includegraphics[width=3.3truein]{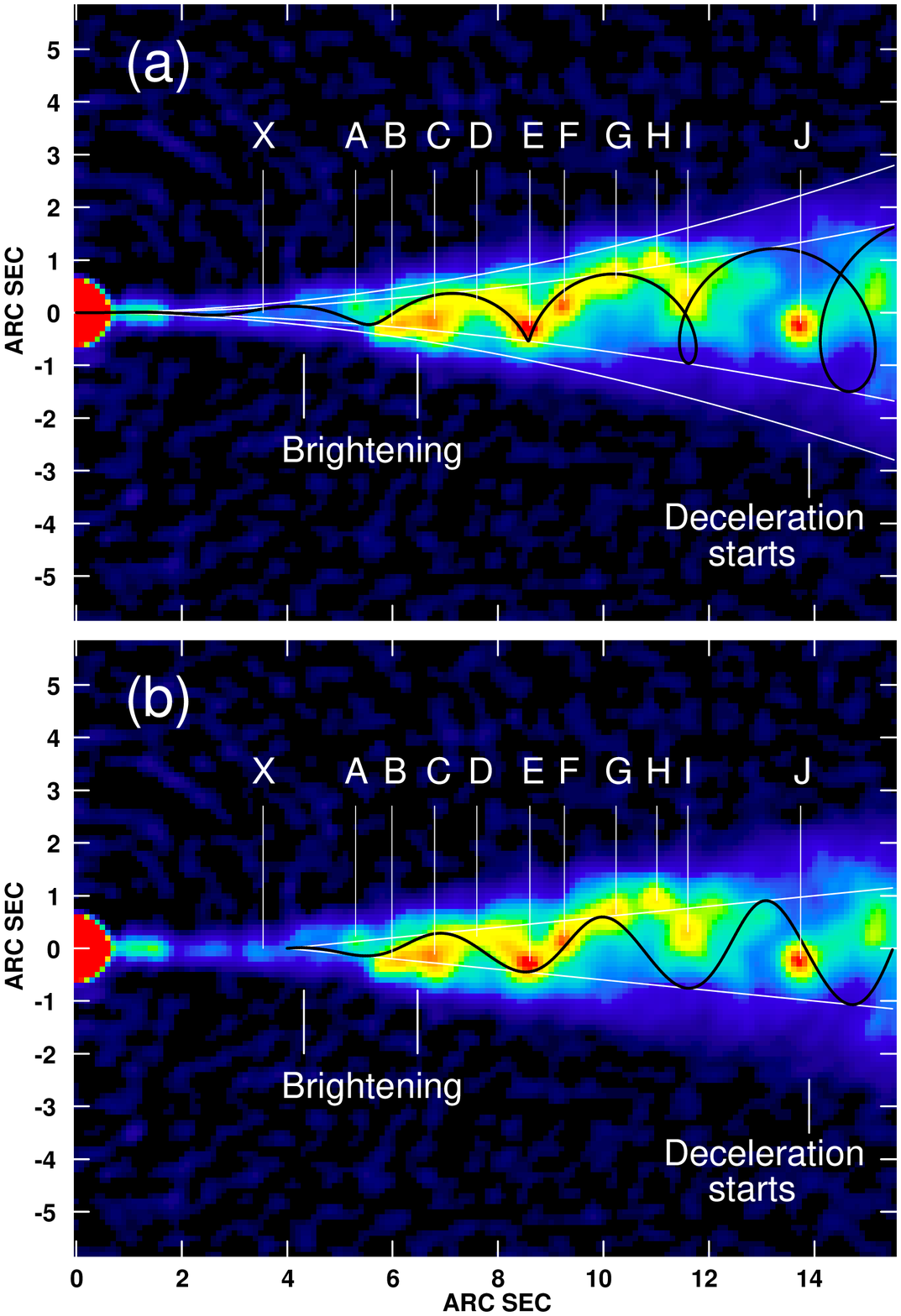}
\caption{False-colour images of the 5-GHz emission of the main jet of
\source\ at a resolution of 0.4~arcsec FWHM, with simple models of the
filament superposed. The locations where the radio jet brightens
abruptly and the start of rapid deceleration from the model of
\citet{canvin} are marked.
`X' marks the location of the inner X-ray knot discussed in
Section~\ref{sec:morp}, and the radio knots of Figure~\ref{fig:radiopolarization} are
also indicated.
(a). The black curve traces a spiral with a projected wavelength of
3.1~arcsec
 lying along the surface at 60 per cent of the width of the
jet envelope. The white curves show the projection of this surface and
the outer boundary for the models of \citet{canvin}.  (b). The black
curve shows the locus of observed positions for components ejected
with a speed of $0.8 c$, as described in the
text, and the white curves show the projected maximum width reached by
the componsnts.  The adopted jet inclination is 38 degrees for both curves.
\label{fig:ridgeline}}
\end{figure}

The radio filament lies wholly within the overall envelope of the
diffuse jet emission, reaching about 60 per cent of the way from the
axis to the edge in projection (Fig.~\ref{fig:ridgeline}a). It is
possible that the filament appears to cross the jet axis only in
projection, and that all of its emission comes from intermediate radii
in the jet.  If so, a promising explanation for the filament's
enhanced emissivity in both the X-ray and radio may be that it lies in
a shear layer between faster inner jet plasma and slower outer
flow. The development of such a shear layer as a jet propagates would
occur at a rate that depends on the rate of mass entrainment through
the boundary layer.  The model transverse velocity-profiles fitted to
this jet by \citet{canvin} have a ratio of edge to on-axis velocity of
$0.8 \pm 0.2$ in the region of interest. Given that the on-axis
velocity is estimated to be $0.9c$, the best-fitting model implies a
large velocity shear, although a top-hat profile is also consistent
with the data.  There is also evidence from much larger distances
\citep[fig.~7 of][]{canvin} where the jet is well resolved, that the
truncated Gaussian functional form assumed for the transverse velocity
profile may under-represent the maximum shear, and that much of the
velocity gradient further down the jet in fact occurs about half way
out from the jet axis. If a similar velocity profile applies closer to
the nucleus, the maximum velocity shear could indeed occur roughly
where we detect the radio and X-ray filament.

The filament could be a magnetic structure, sustained by the velocity
shear, that becomes bright where it is exposed to the same
relativistic electron distribution that causes the brightening of the
diffuse radio emission in the jet.  This is consistent with the
filament representing a similar fraction of the jet emission in both
the radio and X-ray (Sec.~\ref{sec:jetxpec}) and is also consistent with the strong
longitudinal polarization seen in the filament (Sec.~\ref{sec:image}),
but theoretical work is required to test its viability. A specific
mechanism --- field-amplification by the dynamo action of the
turbulence in the shear layer --- is described by \citet{urpin}. In
that case, the filament would be a basically longitudinal feature,
distorted by motions within the shear layer.  The seed field might
then either be internal to the jet or a magnetic structure in the
interstellar medium of \source, entrained into the shear layer
where the jet brightens.

\subsection{The filament as a helical structure?}
\label{sec:helix}

If the filament is a coherent, quasi-helical structure in three
dimensions, rather than a superposition of random, non-axisymmetric
knots which merely appears helical in projection, then we might
constrain its properties further.

The jet envelope is relatively symmetric about the jet axis.
Within this, the appearance of the filament suggests
a series of knots in a
helically-wound structure with roughly two periods of oscillation
across the axis (from A to D, then from D to the region FGH;
Fig.~\ref{fig:ridgeline}).  The oscillation amplitude is a roughly
constant fraction of the jet width.  The filament can be traced as far
as knot I, after which the structure becomes less ordered. It is
tempting to interpret knot E as a cusp where the filament is close to
the line of sight. The black curve in Fig.~\ref{fig:ridgeline}a shows
the projection of a spiral curve whose radius is 0.6 of the jet width
and whose phase has been adjusted to form such a cusp at knot E; the
wavelength along the jet is 1.7~kpc, which projects to 3.1~arcsec on
the plane of the sky.

A simple projection of this sort is only appropriate for a stationary
pattern. The flow speed must be relativistic, in order to avoid seeing
comparably bright structures in the counter-jets of this and other FRI
sources with non-axisymmetric knot structures. If the pattern (as well
as the flow) moves relativistically, then aberration will change its
observed shape.  To quantify this effect we calculated the locus
traced by blobs of emission launched from a precessing injector placed
4~arcsec in projection from the nucleus along the jet axis and
travelling ballistically as in models of SS\,433
\citep{hjellming}. The black curve in Fig.~\ref{fig:ridgeline}b shows
an example for a precession angle of 3.5~deg, a velocity of $0.8 c$,
and a period of 2500 yr. \footnote{Light-travel-time effects cause the
velocity projected on the plane of the sky to be
$v\sin\theta/[1-(v/c)\cos\theta]$ for velocity $v$ along the jet
axis.} The curve is close to sinusoidal with no cusp. Although this is
a toy model taking into account only the kinematics,
the lack of a cusp is a robust result
as the moving pattern is observed close to edge-on in its rest frame
($v/c \approx \cos\theta$). The curve does not match well the
appearance of the filament, making it difficult to understand the
filament as a coherent structure moving with the underlying
relativistic flow.  Interpreting the \source\ filament as a
three-dimensional helix-like structure is therefore problematic
despite its apparently oscillatory appearance.

The filament may have some similarities with the kpc-scale jet in M~87
\citep{owen}, where \citet*{lobanov} model alternations of bright
features from side to side as the superposition of two filaments, with
oscillation wavelengths $\sim0.4$~kpc and amplitudes of order the
width of the radio jet.  A low contrast, two-stranded, filamentary
structure has also been seen in the radio and optical in 3C~66B
\citep{ jackson93}, but no corresponding X-ray structure was found by
\citet{hard01}.

Quasi-periodic oscillatory structures have been seen on parsec scales
in other jets (e.g., \citealt*{lobzen,hardee2}), although it is
unclear whether they are narrow compared with the jet width, as in the
\source\ filament.  One interpretation of these structures (e.g.,
\citealt{abraham,tateyama}) has been that they result from the
ballistic motions associated with precession of the injection
direction when individual knots travel on straight-line trajectories,
as in the simple model discussed above.  In this case, the oscillation
should be strictly periodic and affect the whole jet envelope, which
is not the case in the kpc-scale jet of \source.  Alternatively, the
oscillations have been interpreted as helical gas streaming motions,
perhaps associated with small variations in the injection direction
amplified by Kelvin-Helmholtz instabilities (e.g., \citealt{savolainen}).

\subsection{A jet instability?}
\label{sec:instability}

\subsubsection{Synchrotron thermal instability}

The synchrotron thermal instability \citep{simon,eilek,dalpino} is a
local instability occurring in a magnetized plasma with internal
energy and inertia dominated by relativistic and cold components,
respectively, and has been suggested as a possible cause of
filamentation in radio sources.  \citet{dalpino} show, however, that a
filament formed by this instability would appear darker than its
surroundings in the X rays and brighter at lower (e.g.\ radio)
frequencies.  The brightness contrast in \source\ appears to be
similar in the two wavebands, so the synchrotron thermal instability
is an unlikely cause for the filament.

\subsubsection{Kelvin-Helmholtz instabilities}

Alternatively, we might seek to explain the filament in \source\ as a
hydrodynamic, or magnetohydrodynamic instability if we assume that its
brightness increases with the overpressure generated by the mode
\citep[e.g.,][]{birk,bkh,hardee}. One possibility is that the
instabilities are associated with the jet as a whole.  Alternatively,
the filament may lie within the shear layer between the fast central
and slow outer parts of the jet (Sec.~\ref{sec:filament-nature}). For
a sheared flow, \citet{bkh} showed that typical unstable helical, or
higher-order, Kelvin-Helmholtz modes have narrow pressure maxima
within the shear layer. The thickness of the filament across the jet
might then indicate the width of the shear layer.

While jet distortions are commonly interpreted as Kelvin-Helmholtz
instabilities, such instabilities acting by themselves would not be
expected to generate simple structures because multiple instability
modes are always present together, with similar growth rates, so a
complicated flow pattern would develop. If we regard the filament as a
coherent, quasi-helical structure, then a pure mode or simple mix of
modes is required. Fine tuning would then be needed, for example by
injecting the correct mode by precession of the jet direction, and
then amplifying it by the Kelvin-Helmholtz process, while suppressing
many other fast instabilities \citep[e.g.,][]{birk2002}.  This
problem is alleviated if the filament is a random superposition of
non-axisymmetric knots rather than a coherent structure.

The filament appears to have the symmetry of an $n = 1$
Kelvin-Helmholtz instability mode (where $n$ is the azimuthal mode
number).  A {\em body} (reflection) mode of this type would correspond
to a bulk helical motion of the entire jet. Where the whole structure
is affected, for example the VLBI jet of 3C\,120, this interpretation
is appropriate \citep{hardee2}, but in \source\ the outer envelope of
the jet appears to be undisturbed.  In contrast, a superposition of
$n=1$ and $n=2$ {\it surface} modes was used to fit the double
filament in M\,87 \citep{lobanov}. If the filament in \source\ is a
helical structure, then it must lie well within the volume of the jet,
and is unlikely to be generated by instabilities on the surface.

If the instability is instead restricted to a shear layer, the
azimuthal extent of the filament must still be defined by the
superposition of modes with similar saturated amplitudes. While this
can be achieved by a suitable superposition of (say) $n=1$ and $n=3$,
or $n=2$ and $n=3$, modes, it is hard to see how these modes are
excited while other surface or body modes, which have higher growth
rates, are suppressed.

\subsection{An injection model?}
\label{sec:injection}

We now consider the possibility that the filament is a structure
formed by injection of magnetic field or particles into the shear
layer, either (a) where the injector is stationary and the beam is
rotating about its flow axis, so that the material of the filament is
itself rotating about that axis, or (b) where the injector is rotating
and the beam is non-rotating, so that the filament is being advected
along the beam.

In the case of (a), the observation that the wavelength of the
filament in projection is significantly larger than the half-width of
the jet implies that the rotation velocity of the beam material at the
edge of the jet must be comparable with the component along the axis,
which we infer to be $0.7 - 0.9 c$. Doppler beaming for such a fast
rotation would cause a strong systematic variation in the brightness
of both the filament and the surrounding diffuse emission to either
side of the jet centre-line. This is not seen and we therefore infer
that jet rotation, while expected in principle from any jet-launching
mechanism associated with a rotating structure such as an accretion
disc, is insufficient to shape the filament.

In the case of (b), the velocities of the individual knots in the
filament are outwards along the jet without large rotational
components. The appropriate period is $\sim 2500$~yr, as derived for
the simple ballistic model plotted in Fig.~\ref{fig:ridgeline}.  If
the filament is injected where the jet first brightens, the radius of
150~pc leads to an injection speed comparable with $c$, which is not
reasonable.  If, instead, we assume injection close to the base of the
jet, then the rotational velocities required are more reasonable.  For
a black hole mass of $\sim 2 \times 10^9$ M$_\odot$
(Sec.~\ref{sec:corexpec}), matter in Keplerian motion around the black
hole with a period of 2500~yr would be at a radius of $\sim 1.4$~pc,
which is too large to be within the accretion disk. While gas and
magnetic field from a rotating source, possibly a torus responsible
for X-ray absorption and far-infrared emission
(Sec.~\ref{sec:corexpec}), might be stripped off into the edge of the
jet near its base, this would require that a structure introduced in
the jet on pc~scales survives to the $\sim 5$~kpc scale of the
filament, which is difficult to imagine.

\section{Summary}
\label{sec:summary}

Our \chandra\ observation of \source\ has resolved X-ray jet emission
between 3.6 and 30 arcsec from the core, and also transverse to the
jet axis.  The X-ray spectrum fits a power law with $\alpha_x = 1.2
\pm 0.2$, which is significantly softer than seen in the core, and the
steepening from $\alpha_{\rm r} = 0.61$ through $\alpha_{\rm rx} =
0.89$ to $\alpha_{\rm x} = 1.2$ supports a synchrotron origin for the
jet X-rays. In common with other nearby jets measured with high linear
resolution, the X-ray emission turns on closer to the core than the
position where the radio jet brightens significantly and flares.

The X-ray emission is well matched to the radio in spatial extent
transverse to the jet axis.  The synchrotron nature of the X-rays
requires a mechanism for distributed electron acceleration to TeV
energies.  This mechanism becomes less effective about 14 arcsec from
the core, where the kinematic model for the radio jet derived from
deep radio imaging (Canvin et al 2005) shows that the jet is
decelerating: in this region the ratio of X-ray to radio flux is lower
by a factor of $\sim 3$.

Between the positions of jet brightening and deceleration, a knotty
filament contributes roughly 10 per cent of the emission at both radio
and X-ray wavelengths.  The filament has an oscillatory appearance,
and our radio polarization data show that the magnetic field is
roughly parallel to this filament for most of its length.  The
kinematic model for the jet further from the nucleus requires slower
flow towards the edge of the jet than on the axis, so it is reasonable
to assume that the flow is sheared also in the region of the filament,
where such measurements are not possible from current data.  We
suggest that the filament is a magnetic strand within such a shear
layer. Individual knots in the filament may be somewhat overpressured
with respect to the external X-ray-emitting gas even if they are at
minimum energy.

The geometry of the filament is uncertain. One possibility is that it
is a quasi-helical structure wrapped around the jet axis, although it
is difficult to model the observed shape well if the pattern is moving
relativistically. Alternatively, its appearance as a coherent
structure may be due to a chance superposition of non-axisymmetric
knots or it could be a surface feature whose apparent location inside
the jet volume results from projection.

None of the explanations we have considered for the enhanced emission
from the filament is completely satisfactory:
\begin{enumerate}
\item The synchrotron thermal
instability causes rapid energy loss at high electron energies, such
that the
filament should not be X-ray bright.
\item An explanation in terms of Kelvin-Helmholtz instability modes
requires a
selective superposition of low-order modes, while others with higher
growth
rates are suppressed.
\item There is insufficient surface-brightness asymmetry for the
  emission to arise from material advected along a rotating beam.
\item The origin of the seed magnetic structure is uncertain if the
  filament  is a longitudinal feature on the face of the jet,
  amplified by the dynamo action of turbulence in the shear layer
  and distorted by motions.
\item Injection of particles and/or field from a rotating source 
  close to the brightening point of the jet requires an unreasonably 
  high rotational velocity.
\item Close to the base of the jet, the required rotation period
  for injected field or particles corresponds to a radius of about
  1.4~pc for material rotating around a $\sim 2 \times 10^9$ M$_\odot$ 
  black hole. The problem then is to understand
  how the field or particles might propagate from pc to kpc distances.
\end{enumerate}
Better information on the brightness structure and polarization of the
filament, such as might be obtained by higher-resolution,
higher-sensitivity radio imaging, would test the idea that the
filament represents a well-ordered magnetic structure.  Radio
spectral-index mapping of the filament would test for variations in
the electron energy spectrum, such as might arise from distinct
acceleration regions (knots) within the filament.  Much deeper X-ray
imaging would have a similar purpose, and would allow a more detailed
structural comparison of the radio and X-ray emission of the filament.
Finally, low-surface-brightness non-axisymmetric knottiness is seen in
other FRI radio jets mapped at lower resolution or sensitivity than
\source, and higher-resolution and more sensitive data are required
before we can comment further on the ubiquity or otherwise of coherent
filaments.

The X-ray data for the core have improved understanding of its central
structure.  We have demonstrated that the X-ray data do not allow
sufficient luminosity to be present (but absorbed) to be consistent
with emission close to the Eddington luminosity, suggesting that the central
engine is radiatively inefficient.  The existence of a significant gas
torus is not ruled out by the data since the X-ray emission detected
from the core can be associated with non-thermal radiation from the
small-scale jet rather than emission associated with the accretion
structure.

The X-ray-emitting atmosphere around \source\ has been measured with
the highest precision to date using the \chandra\ data reported
here. Out to 40 arcsec (13.4 kpc), the atmosphere can be described by
a $\beta$ model of core radius $1.7 \pm 0.2$ arcsec and $\beta = 0.52
\pm 0.01$ ($1\sigma$ uncertainties for two interesting parameters),
but at larger radii the intensity drops more sharply.  Significant
temperature structure is measured, ranging from $0.44^{+0.08}_{-0.04}$ keV
close to the radio core, $0.68 \pm 0.03$ keV between 4 and 20 arcsec
from the core, and $0.86^{+0.14}_{-0.06}$ keV beyond 40 arcsec from
the core.  This is consistent with relatively fast cooling times
(e.g., $2.8 \times 10^9$ years at 15 arcsec from the core) that
increase towards the core where the gas density is higher.

\section*{Acknowledgments}

We thank the CXC for its support of \chandra\ observations,
calibrations, data processing and analysis, the SAO R\&D group for
{\sc DS9} and {\sc funtools}, Alexey Vikhlinin for a copy of his
{\sc zhtools} software that has been used for part of our analysis,
and the referee,
John Wardle, for helpful suggestions on improving the clarity and
accessibility of the paper.
We also thank Gijs Verdoes Kleijn for providing galaxy-subtracted HST
images of \source\ and for helpful discussions.  This work has used data
from the VLA.  NRAO is a facility of the National Science Foundation
operated under cooperative agreement by Associated Universities, Inc.

\label{lastpage}

\end{document}